\documentclass[aps,pra,twocolumn,preprintnumbers,amsmath,amssymb,showpacs]{revtex4}
\usepackage{graphicx}
\usepackage{amssymb}
\usepackage{amsmath}
\usepackage{color}
\usepackage[normalem]{ulem}
\usepackage[usenames,dvipsnames]{xcolor}
\usepackage{hyperref}

\newcommand{\eq}{\begin{eqnarray}}
\newcommand{\en}{\end{eqnarray}}
\newcommand{\ket}[1]{|{#1}\rangle}

\begin{document}

\title{Entanglement and the Born-Oppenheimer approximation in an exactly solvable quantum many-body system}

\author{P.A. Bouvrie$^{a}$, A.P. Majtey$^{a,b}$, M.C. Tichy$^{c}$, J.S. Dehesa$^{a}$, A.R. Plastino$^{a,d}$}
\affiliation{
$^a$Instituto Carlos I de F\'{\i}sica Te\'orica y Computacional and
Departamento de F\'{\i}sica At\'omica, Molecular y Nuclear, Universidad de Granada, 18071-Granada, Spain\\
$^b$ Instituto de F\'{i}sica, Universidade Federal do Rio de Janeiro, 21941-972, Rio de Janeiro (RJ) Brazil\\
$^c$ Department of Physics and Astronomy, University of Aarhus, DK-8000 Aarhus C, Denmark \\
$^d$ CeBio y Secretar\'{i}a de Investigaciones, Universidad Nacional
del Noroeste de la Prov. de Buenos Aires, UNNOBA-Conicet,
Roque Saenz-Peña 456, Junin, Argentina\\
%$^{*}$(corresponding author arplastino@ugr.es)
}


\date{\today}

\begin{abstract}

We investigate the correlations between different bipartitions of an exactly solvable one-dimensional many-body Moshinsky model consisting of $N_n$ ``nuclei'' and $N_e$ ``electrons.'' We study the dependence of entanglement on the inter-particle interaction strength, on the number of particles, and on the particle masses. Consistent with kinematic intuition, the entanglement between two subsystems vanishes when the subsystems have very different masses, while it attains its maximal value for subsystems of comparable mass. We show how this entanglement feature can be inferred by means of the Born-Oppenheimer Ansatz, whose validity and breakdown can be understood from a quantum information point of view.

\end{abstract}

\maketitle

\section{Introduction}

The very definition of entanglement relies on the partitioning of a system into subsystems, such that one physical system can exhibit very different entanglement properties, depending on the assumed convention \cite{TichyMintert2011}.
Efficiently solvable systems permit one to find a particular partition for which the wavefunction is separable, even in the presence of otherwise entangling interaction. For example, the hydrogen atom is naturally treated in the center-of-mass and relative coordinates, in which the wavefunction factorizes, instead of in the electron and proton coordinates, in which the wavefunction appears to be highly entangled \cite{TommasiniTimmermansEtal1998}. Such a beneficial change of coordinates is, however, impossible for non-integrable systems, and entanglement in quantum many-body systems typically occurs with unconquerable complexity, representing a serious challenge to any numerical or analytical approach. By definition, in a quantum chaotic system, there is no basis to the Hilbert-space that permits an efficient description. Quantities developed in quantum information reflect the failure of any strategy that relies on the truncation of the Hilbert space, {\it e.g.}
by the statistics of Schmidt coefficients of the wavefunction described under any bipartition \cite{VenzlDaleyEtal2009}. Conversely, fundamental restrictions on entanglement, {\it e.g.}~by area laws \cite{EisertCramerPlenio}, can render efficient simulations of quantum-many-body systems possible \cite{PerezGarciaVerstraeteEtal2007}. An understanding of entanglement can, thus, be of great importance for practical numerical solutions.

A system that is particularly prone to complexity is the many-electron atom, in which the long-range Coulomb interaction renders any exact solution impossible, as with helium \cite{TannerRichter2000}. For such a system, entanglement is representative of the enormous complexity present in the system.

Indeed, the main features that were found in the analytical treatment of simplified models of helium-like systems \cite{YanezPlastinoEtal2010,ManzanoPlastinoEtal2010} also persist in results based on numerical studies with high-quality wavefunctions \cite{DehesaKogaEtal2012,BenentiSiccardiEtal2013}: entanglement between electrons tends to increase with the eigenenergy.
This increase is also observed in the case of the singlet-states of helium, but not for triplet-states \cite{DehesaKogaEtal2012}, for which no satisfactory explanation is yet available.
Entanglement also increases, in general, with the interaction strength between constituents \cite{ManzanoPlastinoEtal2010}, which is consistent with the decrease in correlations experienced when a strong external field shields inter-particle interactions due to confinement \cite{BouvrieMajteyEtal2012}. Another property exhibited by the atomic systems studied so far is that the entanglement of excited states does not necessarily vanish in the limit of weak interactions \cite{YanezPlastinoEtal2010}. This feature, as well as the tendency of entanglement to increase with the eigenenergy, has been shown to be closely related to the degeneracy of the energy levels of the associated independent particle model obtained in the limit of vanishing interaction \cite{MajteyPlastino2012}. Both properties will be revealed throughout the paper.

An analytically solvable model that can be applied to virtually any number of particles is the Moshinsky atom \cite{Moshinsky1968,MoshinskyMendezEtal1985} (sometimes referred to as ``harmonium'' \cite{BenavidesGarcia2012}), for which all appearing potentials are set to be harmonic. The application of harmonium as a tractable model for elucidating some aspects of the behaviour of more
realistic systems has a long history which goes back almost to the very beginning of quantum mechanics \cite{Heisemberg1926}. This model allows an analytical solution, and therefore constitutes a valuable testing bench for the study of diverse aspects of atomic and molecular physics. Indeed, it has been used for assessing the quality of the Hartree-Fock approximation \cite{Moshinsky1968} and several density functionals \cite{BenavidesVarilly2012}, but also for investigating low-order density matrix descriptions of the ground state of atomic systems \cite{AmovilliMarch2003}, and for exploring entanglement-related features \cite{AmovilliMarch2004,PipekNagy2009,BenavidesGarcia2012,YanezPlastinoEtal2010,BouvrieMajteyEtal2012} and other manifestations of quantum correlations \cite{LagunaSagar2011} in atomic systems and in many distinguishable particles \cite{KoscikOkopinska2013}.
This model has also been found useful in the study of other subjects beyond atomic physics, such as the thermodynamics of black holes \cite{Srednicki1993}.

In this paper we deal with a many-particle harmonic model with different masses to simulate a ``molecule'' with an arbitrary number of nuclei and electrons in an external harmonic potential. The clear hierarchy within the masses of the molecular constituents suggests that most of the entanglement properties can already be understood from purely kinematical considerations.
A well-established computational technique in physical chemistry is the Born-Oppenheimer approximation, which allows an efficient treatment of systems with many nuclei and many electrons thanks to the particle-mass scale.
The validity and scope of the Born-Oppenheimer approximation has been studied and tested for different systems, {\it e.g.} for harmonic models \cite{cederbaum2013}, for atoms \cite{takahashiKazuo2006} and molecules in magnetic fields \cite{SchmelcherCederbaumEtal1988}, or in chemical reactions \cite{AlexanderCapecchiEtal2002} and in nonadiabatic theories \cite{YoneharaHanasakiEtal2012}.

Here, we show that the entanglement present in many-particle systems can be understood to range widely not only from purely kinematic considerations, but also from the Born-Oppenheimer Ansatz, which
permits us to assess the validity of the approximation itself in zeroth adiabatic electron theories.
We also investigate the entanglement properties of the eigenstates of this many-particle Moshinsky-like model for different bipartitions of the system through the parameters that characterize it; namely the strength of the interactions between particles, the number of particles, and their corresponding masses.

We first describe the exactly solvable many-particle Moshinsky model in section \ref{Sec:ExactSolution}. In section \ref{Sec:PrelimEntMeasure}, we briefly review the entanglement measures to be used and then show their properties for the particular case of three-particle in section~\ref{Sec:3PaExactModel}. We extend the study of entanglement to systems with an arbitrary number of particles in section \ref{Sec:NPaExactModel}, and in \ref{Sec:BOManyParticleModel} we deal with the Born-Oppenheimer approximation for this many-particle Moshinsky model. Finally, some conclusions are drawn in section \ref{Sec:Conclusions}.

\section{The many particle system}
\label{Sec:ExactSolution}

We consider a system of $N=N_n+N_e$ distinguishable particles, consisting if $N_n$ ``nuclei'' with mass $m_n$ and $N_e$ ``electrons'' with mass $m_e$. All particles interact harmonically with each other and with the external confining potential.

The interparticle-interaction strengths between a nucleus and an electron, between two electrons and between two nuclei are denoted by $\tau_{ne}$, $\tau_{ee}$ and $\tau_{nn}$, respectively; they are measured in units of the confining potential strength $k$.
All masses are measured in units of the electron mass $m_e$, {\it i.e.}~the nucleus mass is adjusted via the mass ratio $M=m_n/m_e$, and all actions are measured in units of $\hbar$.

It is worth mentioning that throughout this work we consider that electrons are distinguishable and do not carry spin. Taking into account the indistinguishability of particles could add new qualitative features as compared to the model of distinguishable particles \cite{YanezPlastinoEtal2010,BouvrieMajteyEtal2012,MajteyPlastino2012}.

Since our model is separable between the three dimensions, then,
and in order to simplify notation without causing any loss of significant physical results, we consider a 
one-dimensional many-body system.
 The dimensionless Hamiltonian of the system is
\eq
\label{ManyParticleHamiltonian}
H_x = \sum _{i=1}^{N_n} \frac{{P_{x_i}}^2}{2M}+\sum _{j=1}^{N_e} \frac{{p_{x_j}}^2}{2}  +
\frac{1}{2} \sum _{i=1}^{N_n} {X_i}^2+ \frac{1}{2} \sum _{j=1}^{N_e} {x_j}^2 + \nonumber \\
+\frac{\tau_{ne}}{2}\sum _{i=1}^{N_n} \sum _{j=1}^{N_e} (X_i-x_j)^2 +
\frac{\tau_{ee}}{2}\sum _{j=1}^{N_e} \sum _{k=j+1}^{N_e} (x_j-x_k)^2 +  \nonumber \\
+\frac{\tau_{nn}}{2}\sum _{i=1}^{N_n} \sum _{k=i+1}^{N_n} (X_i-X_k)^2
 ,
\en
where $X_j,P_{x_j}$ and $x_j, p_{x_j}$ denote the positions and momenta of the nuclei (in uppercase letters) and electrons (in lowercase letters), respectively.

Details on the derivation of the exact eigenfunctions and eigenenergies of the Hamiltonian  \eqref{ManyParticleHamiltonian} are located in Appendix \ref{Sec:AppendixA}.

The coordinates that allow us to rewrite the system Hamiltonian $H_x$ in a fully separable form are given by $N_n-1$ and $N_e-1$ Jacobi relative variables for nuclei \eqref{JacobiChangeRelative2} and electrons \eqref{JacobiChangeRelative1} respectively, besides the two coordinates
\eq
\label{ChangeU1}
U_1 (R_{N_n},r_{N_e}) &=& \frac{N_n (a+b) R_{N_n}+N_e r_{N_e}}{\sqrt{N_e+N_n (a+b)^2}}, \\
\label{ChangeU2}
U_2 (R_{N_n},r_{N_e}) &=& \frac{N_n (a-b) R_{N_n}+N_e r_{N_e}}{\sqrt{N_e+N_n (a-b)^2}},
\en
where $R_{N_n}$ and $r_{N_e}$ are the center-of-mass coordinates for the nuclei and electrons, respectively. The parameters $a$ and $b$, given in Eq.~\eqref{ChangeParameters}, are functions of $\tau_{ne}$ and $M$.

All correlations between nuclei and electrons are encoded in the correlations between their respective centers of mass $R_{N_n}$ and $r_{N_e}$, which are coupled only through the coordinates $U_1$ and $U_2$. In the limit of large electron-nucleus interaction $\tau_{ne}\rightarrow\infty$, these coordinates become the center-of-mass and the relative coordinates of the set of nuclei and electrons
\eq
\label{eq:ChangeULimitIntau}
\lim_{\tau_{ne}\rightarrow\infty} U_1 &=& \sqrt{N_e+M N_n} \frac{N_e r_{N_e}+M N_n R_{N_n}}{{N_e+M N_n}}, \nonumber \\
\lim_{\tau_{ne}\rightarrow\infty} U_2 &=& \sqrt{\frac{M N_e N_n}{N_e +M N_n}} (r_{N_e}-R_{N_n}).
\en

When electrons and nuclei have the same mass, {\it i.e.}~$M=1$, the coordinates again allow the above natural interpretation
\eq
\label{eq:ChangeUM1}
U_1^{M=1} &=& \sqrt{N_e+ N_n} \frac{N_e r_{N_e}+ N_n R_{N_n}}{{N_e+ N_n}}, \nonumber \\
U_2^{M=1} &=& \sqrt{\frac{N_e N_n}{N_e + N_n}} (r_{N_e}-R_{N_n}) ,
\en
for any value of the interaction $\tau_{ne}$.

In the following, we denote pure states of the system \eqref{ManyParticleHamiltonian} by $|u_1,u_2,n_1,\ldots,n_{N_n-1},e_1,\ldots,e_{N_e-1}\rangle$ where the quantum numbers $u_1, u_2, n_i, e_j$ correspond to the excitation of each collective coordinate $U_1$, $U_2$ and of the $i(j)$th nuclei (electrons) Jacobi coordinate respectively.

\section{Entanglement measure and the selected bipartitions}
\label{Sec:PrelimEntMeasure}

In this paper we focus on the bipartite entanglement in eigenstates of the many-particle Moshinsky-model described above. For this purpose we consider different bipartitions: First, we divide the system into two groups, one containing all $N_n$ nuclei and the other containing all $N_e$ electrons. Second, we study the correlations of a single particle (electron or nucleus) with the rest of the system.

The entanglement of a pure bipartite system is essentially given by the mixedness of the marginal density matrices associated with each subsystem. A practical quantitative indicator for the entanglement in a pure bi-partite system is the linear entropy \cite{AmicoFazioEtal2008}
\begin{equation}\label{entanglementdef}
\varepsilon(|\psi\rangle)=1-\text{Tr}[\rho_{A}^2]=1-\text{Tr}[\rho_{B}^2],
\end{equation}
where $\rho_A$ and $\rho_B$ are the reduced density matrices of subsystems $A$ and $B$, respectively. For separable pure states $\ket{\psi}=\ket{\phi_A}\ket{\phi_B}$, this quantity vanishes. In the present applications, we deal with infinite-dimensional Hilbert spaces, such that the measure (\ref{entanglementdef}) adopts values in the interval $[0,1)$, since the maximal value of the entanglement in a $d$-dimensional space is $\varepsilon_\text{max}(\ket{\psi})=1-1/d$.

The linear entropy has several computational advantages, both analytically and numerically, over other measures, such as the von Neumann entropy, $\varepsilon^{(vN)}(|\psi\rangle) = S[\rho_A] = -\text{Tr}[\rho_{A} \ln \rho_{A}]$.
In particular, and contrary to the von Neumann entropy $S[\rho_A]$, the computation of the linear entropy $1-\text{Tr}[\rho_A^2]$ does not require the diagonalization of the density matrix $\rho_A$. The linear entropy \eqref{entanglementdef} coincides, up to multiplicative and additive constants, with some measures of entanglement monotone \cite{Vidal2000}%\cite{Wootters1998}
, which proved to be a powerful tool for elucidating many aspects of the entanglement properties of pure states (see, for instance, \cite{BuscemiBordoneEtal2007,PlastinoManzanoEtal2009,TichyMintert2011,YanezPlastinoEtal2010,MajteyPlastino2012,BouvrieMajteyEtal2012,TichyBouvrie2012a,TichyBouvrie2012b}).

Since the constituents of atoms, molecules and, in general, many interacting particles systems are usually highly entangled with the rest of the system, any reliable Schmidt representation of the system state has to have a large number of non-negligible independent contributions, {\it i.e.} a large number of non-negligible Schmidt coefficients. This makes difficult \cite{PerezGarciaVerstraeteEtal2007,DehesaKogaEtal2012}, if not impossible \cite{VenzlDaleyEtal2009}, any proper simulation of the state and its quantum correlations. The Moshinsky model however, admits an analytical computation of the infinite Schmidt series, as demonstrated for the ground state in the Ref.~\cite{PruskiEtal1972}, which made possible the entanglement study of the ground state of any bipartition in the many-identical-particle Moshinsky model \cite{KoscikOkopinska2013}.

Here, we compute the entanglement in a continuous variable framework, thus avoiding the intricate diagonalization procedure of the reduced density matrix. The method described below allows us to extend the entanglement studies done in Ref.~\cite{KoscikOkopinska2013} not only to systems with different particle masses, but to excited states as well.

Given a bipartition $(A,B)$ of a system of $N$ particles into $(N_A,N_B)$ particles, we compute the trace that appears in \eqref{entanglementdef} as
\begin{equation}
\label{deftraza}
\text{Tr}[\rho_A^2]=\int_{\mathbb{R}} |\langle {\bf x}_A|\rho_A|{\bf x}'_A\rangle|^2 d{\bf x}_Ad{\bf x}'_A,
\end{equation}
with the matrix elements of $\rho_A$ given by
\eq
\langle {\bf x}_A|\rho_A|{\bf x}'_A \rangle = \int_{\mathbb{R}} \langle {\bf x}_A{\bf x}_B|\rho|{\bf x}'_A {\bf x}_B\rangle d{\bf x}_B = \nonumber \\
\int_{\mathbb{R}} \Psi({\bf x}_A,{\bf x}_B)\Psi^*({\bf x}'_A,{\bf x}_B) d{\bf x}_B,
\en
where ${\bf x}_A$ (${\bf x}_B$) are $N_A$-dimensional ($N_B$-dimensional) position coordinates denoting the global set of coordinates $\{x_1 \cdots x_{N_A}\}$ ($\{x_{N_A+1} \cdots x_{N}\}$) of the particles that belong to subsystem $A$ ($B$).

For our choices of subsystem partitions, we denote
the ($N_n$-nuclei)-($N_e$-electrons) entanglement (or nuclei-electrons entanglement) by $\varepsilon$, the (1-nucleus)-($(N-1)$-particles) entanglement (or nucleus entanglement) by $\varepsilon_n$, and the (1-electron)-($(N-1)$-particles) entanglement (or electron entanglement) by $\varepsilon_e$. The electron (nucleus) entanglement captures the uncertainty that a single electron (nucleus) is subject to due to correlations with other particles. These qualitative correlations can be of a very distinct nature because electrons can be correlated with each other, or with the nuclei. This is also reflected in the nuclei-electrons-entanglement.

These three different types of entanglement ($\varepsilon$, $\varepsilon_n$ and $\varepsilon_e$) depend on the parameters and quantum numbers shown in Table \ref{TableDependences}.

\begin{table}[h]
\begin{tabular}{|c|c|c|c|c|}
\hline
Parameters & $\varepsilon$ & $\varepsilon_n$ & $\varepsilon_e$ \\
\hline
Interactions & $\tau_{ne}$ & $\tau_{ne}$, $\tau_{nn}$   & $\tau_{ne}$, $\tau_{ee}$  \\
Quantum numbers & $u_1$,$u_2$ & $u_1$,$u_2$,$n_{N_n-1}$ & $u_1$,$u_2$,$e_{N_e-1}$ \\
Mass and particles & $M,N_n,N_e$ & $M,N_n,N_e$ & $M,N_n,N_e$ \\
\hline
\end{tabular}
\caption{Parameters on which the different types of entanglement depend.}
\label{TableDependences}
\end{table}

The choice of the coordinate changes \eqref{JacobiChangeRelative1}-\eqref{JacobiChangeCentermass} is especially suited for our study on entanglement, since the nuclei-electrons entanglement depends only on the quantum number associated to $r_{N_e}$ and $R_{N_n}$ and the electron (nucleus) entanglement is independent of the collective excitation of the nuclei (electrons).

Since the interactions between electrons and between nuclei are irrelevant for the nuclei-electrons entanglement $\varepsilon$, one would be lead to infer from Table~\ref{TableDependences} that we can treat the subsystems of nuclei and electrons as two entities with masses $MN_n$ and $N_e$, respectively, which  interact mutually with some effective strength. But surprisingly, this does not happen in general and, consequently, the parameters $M$, $N_n$ and $N_e$ cannot be rescaled between them.

In the limiting case $\tau_{ne}\rightarrow\infty$ \eqref{eq:ChangeULimitIntau} (as well as for $M=1$ \eqref{eq:ChangeUM1}) the change of variables $U_1$ and $U_2$ are precisely the center of mass and the relative coordinates of $R_{N_n}$ and $r_{N_e}$ (the center of mass of the nuclei and electrons, respectively). These are the special cases for which the subsystems can be treated as two single entities and parameters $M$, $N_n$ and $N_e$ can be rescaled as $\gamma=MN_n/N_e$. Otherwise, the coordinates $U_1$ and $U_2$ depends on $R_{N_n}$ and $r_{N_e}$ in a significant way, and the above intuitive reading is wrong.

\section{Entanglement of the three-particle system}
\label{Sec:3PaExactModel}

\begin{figure}[ht]
\includegraphics[scale=0.6]{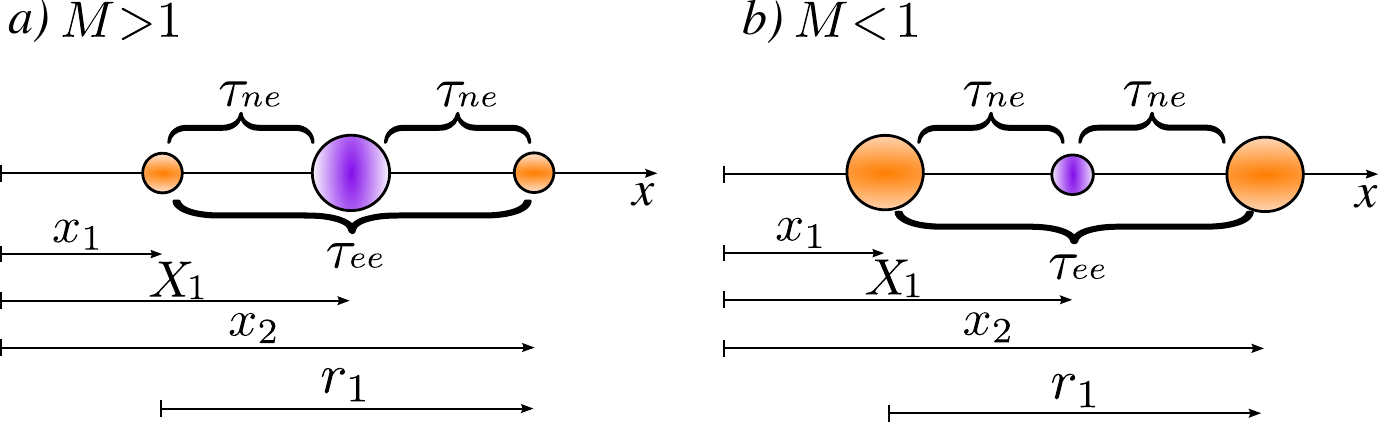}
\caption{(Color online) Three harmonically interacting particles in a confining external harmonic potential, namely $N_n=1$ and $N_e=2$, which roughly describes $a)$ an $H\text{e}$-like system for $M\gg 1$ and $b)$ a $H_2^+$-like system for $M\ll 1$. The sizes of the circles symbolize the masses of the particles.}
\label{ThreeParticlesSystem}
\end{figure}

In the particular case of a three-particle system with one nucleus ($N_n=1$) and two electrons ($N_e=2$), the possibility of choosing different particle mass ratios $M$ allows us to qualitatively model two different physical systems: a helium-like atom for $M \gg 1$, and a diatomic molecule ($H_2^+$ type) with one ``electron'' for $M\ll 1$, as shown in Figs.~\ref{ThreeParticlesSystem}$a)$ and $b)$, respectively. We keep the notation for the nuclei $m_n$ and electrons $m_e$, even though $m_e\gg m_n$ ($M\ll 1$).

In the following subsections we determine analytically the entanglement given by Eq.~\eqref{entanglementdef}, for various low-lying states $|u_1,u_2,e_1\rangle$; namely, the ground state $|000\rangle$ and the first excited states $|100\rangle$, $|010\rangle$ and $|001\rangle$.

\subsection{Level degeneracy and limit of vanishing interaction}

As shown in Refs.~\cite{MajteyPlastino2012,BouvrieMajteyEtal2012}, an infinitesimal inter-particle interaction can give rise to excited states with finite entanglement. The degenerate eigenstates $|\psi_j\rangle$ (all with the same energy) of a non-interacting system $H_0$ can always be chosen to be a separable state (non-entangled). If we solve the eigenvalue problem corresponding to the (perturbed) Hamiltonian
\eq
H=H_0+\tau H'
\en
and take the limit $\tau \to 0$, the perturbation $H'$ will lift the degeneracy at least partially and ``choose'' one particular basis among the infinite possible bases, whose states are generally entangled. Therefore, in the limit of vanishing interaction, the entanglement exhibited by the Hamiltonian \eqref{ManyParticleHamiltonian} is finite for excited states that are degenerate in this limit.

In this subsection we will carry out a similar analysis as performed in \cite{MajteyPlastino2012,BouvrieMajteyEtal2012} but, unlike the model used there, the Hamiltonian $H_x$ will have two different particle species which contribute to reducing the degeneracy of the energy levels. The energy of the state $|u_1,u_2,e_1\rangle$, given by \eqref{eigenenergies}, in the limit of vanishing interactions ($\tau_{ne}\rightarrow 0$, $\tau_{ee}\rightarrow 0$) reads
\eq
E'_0 =
\left\{
\begin{array}{ccc}
\label{energy-limit}
 1+\frac{1}{\sqrt{M}}\left(\frac{1}{2}+u_1\right)+u_2+e_1 & \text{if} & M \geq 1 \\
 1+u_1+\frac{1}{\sqrt{M}}\left(\frac{1}{2}+u_2\right)+e_1 & \text{if} & 0<M<1.
\end{array}
\right.
\en

\begin{figure}[ht]
\includegraphics[scale=0.55]{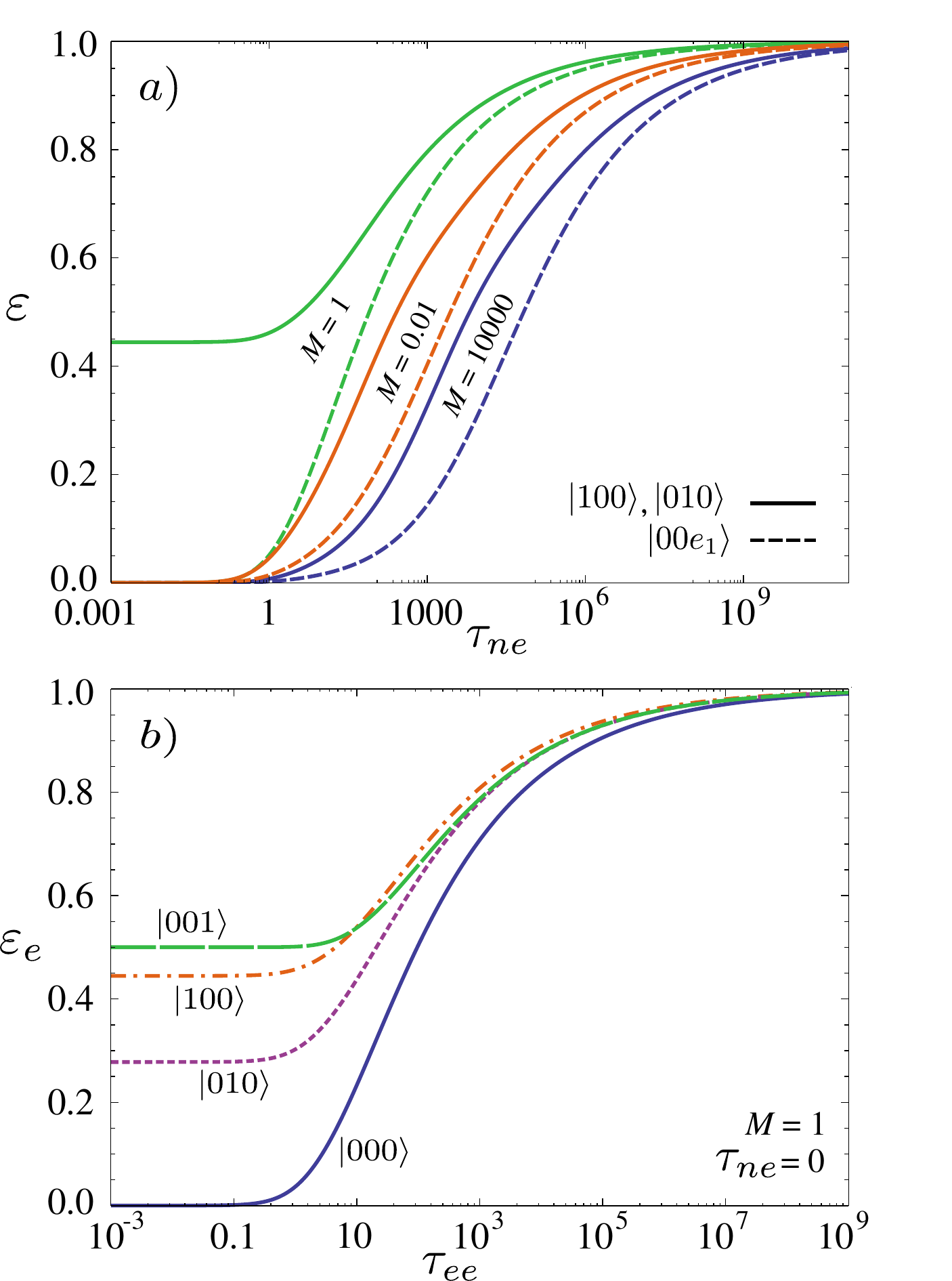}
\caption{(Color online) $a)$ Ground state (dashed lines) and first excited states (solid and dashed lines) nucleus-electrons entanglement, $\varepsilon(\tau_{ne},M)$, of the three-particle case as a function of the nucleus-electron interaction strength $\tau_{ne}$ and for different mass ratio $M$. $b)$ Single-particle entanglement, $\varepsilon_e(\tau_{ne},\tau_{ee},M)$, of a system with three identical particles ($M=1$) as a function of $\tau_{ee}$, with $\tau_{ne}=0$, for the ground state and first few exited states.}
\label{EntNuctau1}
\end{figure}

The nucleus-electron entanglement $\varepsilon$ reflects the correlations between the nucleus and the two electrons. It depends on
the quantum numbers $u_1$ and $u_2$ and the interaction strength $\tau_{ne}$, but it depends neither on the excitation of the electron relative coordinate $e_1$, nor on the interaction $\tau_{ee}$, (see Table~\ref{TableDependences}), {\it i.e.} the nucleus does not feel the inter-electronic structure. Therefore, the nucleus entanglement of the state $|00e_1\rangle$ is less entangled than any excited state in $u_1$ or $u_2$ (see Fig.~\ref{EntNuctau1}$a)$.

The electron entanglement $\varepsilon_e$ reflects the correlations between one electron and the remaining particles of the system, namely the other electron and the nucleus. It depends on both interaction strengths $\tau_{ne}$ and $\tau_{ee}$, and on all quantum numbers, $u_1$, $u_2$ and $e_1$.

As in the general trend shown in Refs.~\cite{YanezPlastinoEtal2010,ManzanoPlastinoEtal2010,BouvrieMajteyEtal2012,KoscikOkopinska2013}, the entanglement increases with the interaction for all states, (see Fig.~\ref{EntNuctau1}). However, in the limit of vanishing interaction $\tau_{ne}\to 0$, the nucleus decouples from the electrons and $\varepsilon$ always vanishes unless the energy levels are degenerate. This is the case for the excited states $|010\rangle$ and $|100\rangle$ when $M=1$, which exhibit a finite amount of entanglement (see Fig.~\ref{EntNuctau1}$a)$.
The energy levels are degenerated in this limit \eqref{energy-limit} when $u_1=u_2$ and $M=1$ (for any $e_1$ of which the entanglement is independent).

In the limit of both vanishing interactions, $\tau_{ne}\rightarrow 0$ and $\tau_{ee}\rightarrow 0$, a finite amount of the electron entanglement $\varepsilon_e$ is observed for the states $|100\rangle$ ($|010\rangle$) if $M<1$ ($M> 1$). From Eq.~\eqref{energy-limit} we note that the energy level of the state $|100\rangle$ ($|010\rangle$) is degenerate when $M< 1$ ($M> 1$), and it has the same energy as $|001\rangle$. The energy level of the state $|001\rangle$ is degenerate for all $M$ values, and a finite amount of entanglement is always obtained in the limit of vanishing interactions. For $M=1$, all excited states have a finite entanglement in these limits, as shown in Fig.~\ref{EntNuctau1}$b)$, due to the degeneracy of the energy when any of the quantum numbers $u_1$, $u_2$ and $e_1$ are equal.

\subsection{Kinematic considerations}

\begin{figure}[ht]
\includegraphics[scale=0.55]{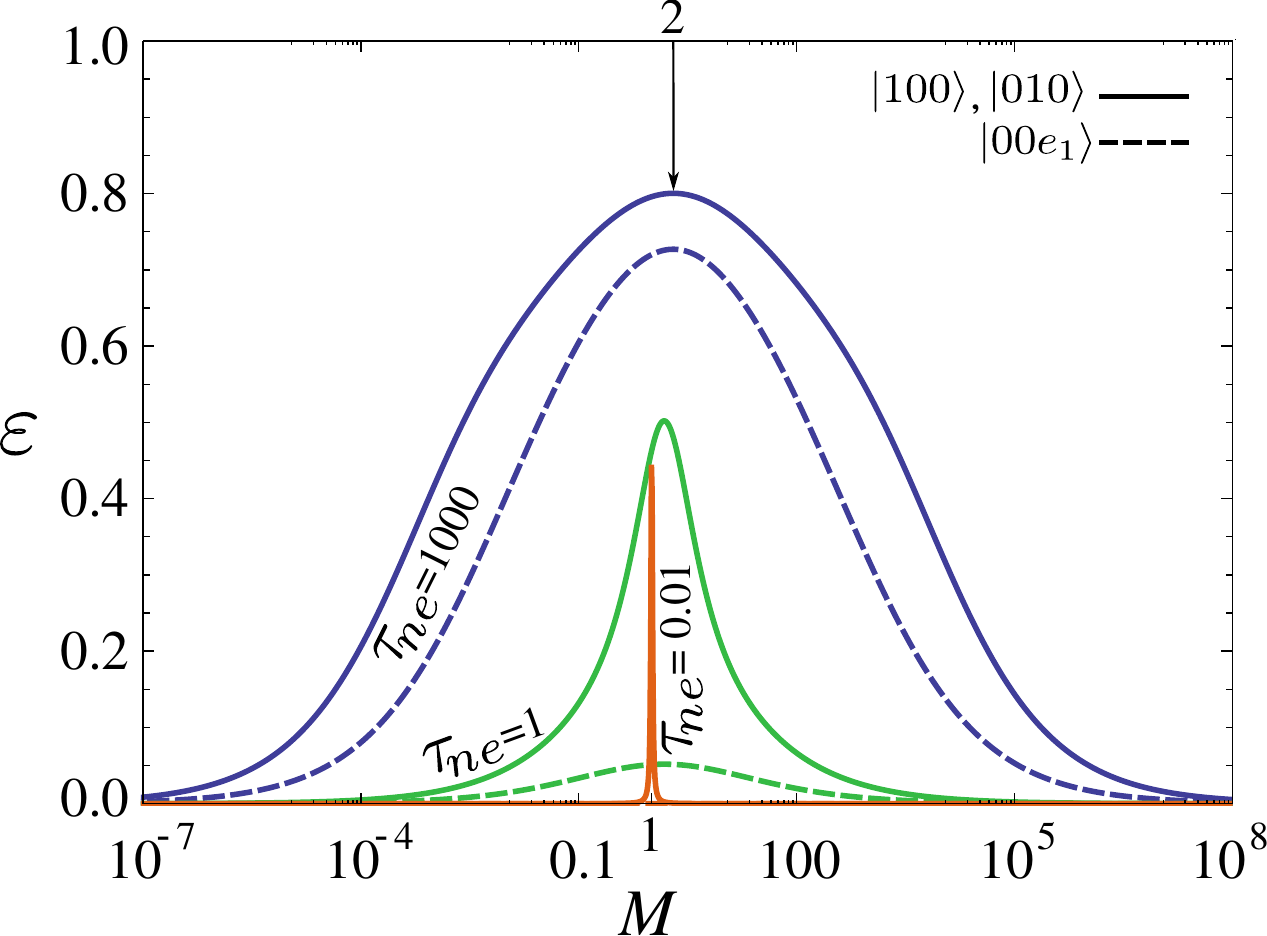}
\caption{(Color online) Nucleus-electrons entanglement, $\varepsilon(\tau_{ne},M)$, in the three-particle case as a function of the mass ratio $M$. We plot the ground state (dashed lines) and first few excited states (solid and dashed lines) for different values of the interaction $\tau_{ne}$.}
\label{EntNucM}
\end{figure}

In the previous subsection we pointed out, as done in Refs.~\cite{MajteyPlastino2012,BouvrieMajteyEtal2012}, that the physics of the system in the limit of vanishing interaction can only be understood within a quantum framework; it is extremely affected by the degeneracy of the energy levels. However, in the case of strongly interacting particles one can appeal to classical kinematic intuition. Thus, states of systems with very different subsystem masses are less entangled than states with similar subsystems masses. This is reflected in Fig.~\ref{EntNuctau1}$a)$, and more evidently in Fig.~\ref{EntNucM} where we plot $\varepsilon$ as a function of the mass ratio $M$ for different values of the interaction $\tau_{ne}$.

When $\tau_{ne}\gg 1$, the maximal entanglement is achieved for $M \simeq 2$, which corresponds to subsystems with the same mass. As the masses become more different the entanglement gradually fades. The more equal the masses two coupled systems have, the more they will influence each other reciprocally. In terms of the physical limits, for a weak interaction, particles are independent, even if they have the same mass (if there is no degeneracy in this limit, see in Fig~\ref{EntNucM} the jump in entanglement for $\tau_{ne}=0.01$). For very large or very small mass ratio, the heavy particles are not influenced much by the light ones and, additionally, the light particles are still in a rather pure state.

At this point, the obvious question that arises is whether this kinematic property persists when the number of particles in the system is higher.

\section{Many-particle entanglement}
\label{Sec:NPaExactModel}

The great advantage of the many-particle model at issue here is that one can determine analytically the entanglement for an arbitrary number of particles of the bipartitions. Using the exact eigenstates of the Hamiltonian \eqref{ManyParticleHamiltonian}, given in Appendix \ref{Sec:AppendixA}, one can compute the integrals involved in Eq.~\ref{deftraza}. In this section we evaluate analytically the ground state entanglement for the three bipartitions $\varepsilon$, $\varepsilon_n$ and $\varepsilon_e$, by means of the linear entropy defined in Eq. \eqref{entanglementdef}. We discuss and argue the main features of the entanglement as a function of the different parameters of this many-particle system, particularly the number of particles. For the different bipartitions the parameters are given in Table~\ref{TableDependences} where we consider $M>1$ in what follows.

\subsection{Nuclei-electrons entanglement}
\label{Sec:ManyPartNucElecEnt}

For non-negligible interaction, $\tau_{ne}$, the nuclei-electrons entanglement, $\varepsilon$, displays a remarkable general trend: it is maximal when the mass ratio $M$ fulfils $M\simeq N_e/N_n$. This behavior can be understood from the coordinate changes $U_1$ and $U_2$ in the limit of large interaction, Eq.~\eqref{eq:ChangeULimitIntau}. The more similar the contributions of $R_{N_n}$ and $r_{N_e}$ are, the larger is the correlation between nuclei and electrons and, hence, the entanglement.

Maximal entanglement is reached exactly at $M = N_e/N_n$ for any finite interaction, $\tau_{ne}$, only when $N_n = N_e$ ($M = 1$). In such a case, nuclei-electrons entanglement describes the correlations between two particles with unequal masses $m_1=MN_n$ and $m_2=N_e$, and with some interaction strength $\tau$. However, this does not happen if the considered subsystems have a different number of particles $N_n \neq N_e$ because the symmetry in the number of interactions per particle in each subsystems is lost.
The parameters $M$, $N_n$ and $N_e$ cannot be rescaled and one cannot consider each subsystem as a single entity. In systems with $N_n \neq N_e$, the maximal entanglement depends on the relative nucleus-electron interaction strength $\tau_{ne}$ and the mass ratio $M$.

\begin{figure}[ht]
\includegraphics[scale=0.6]{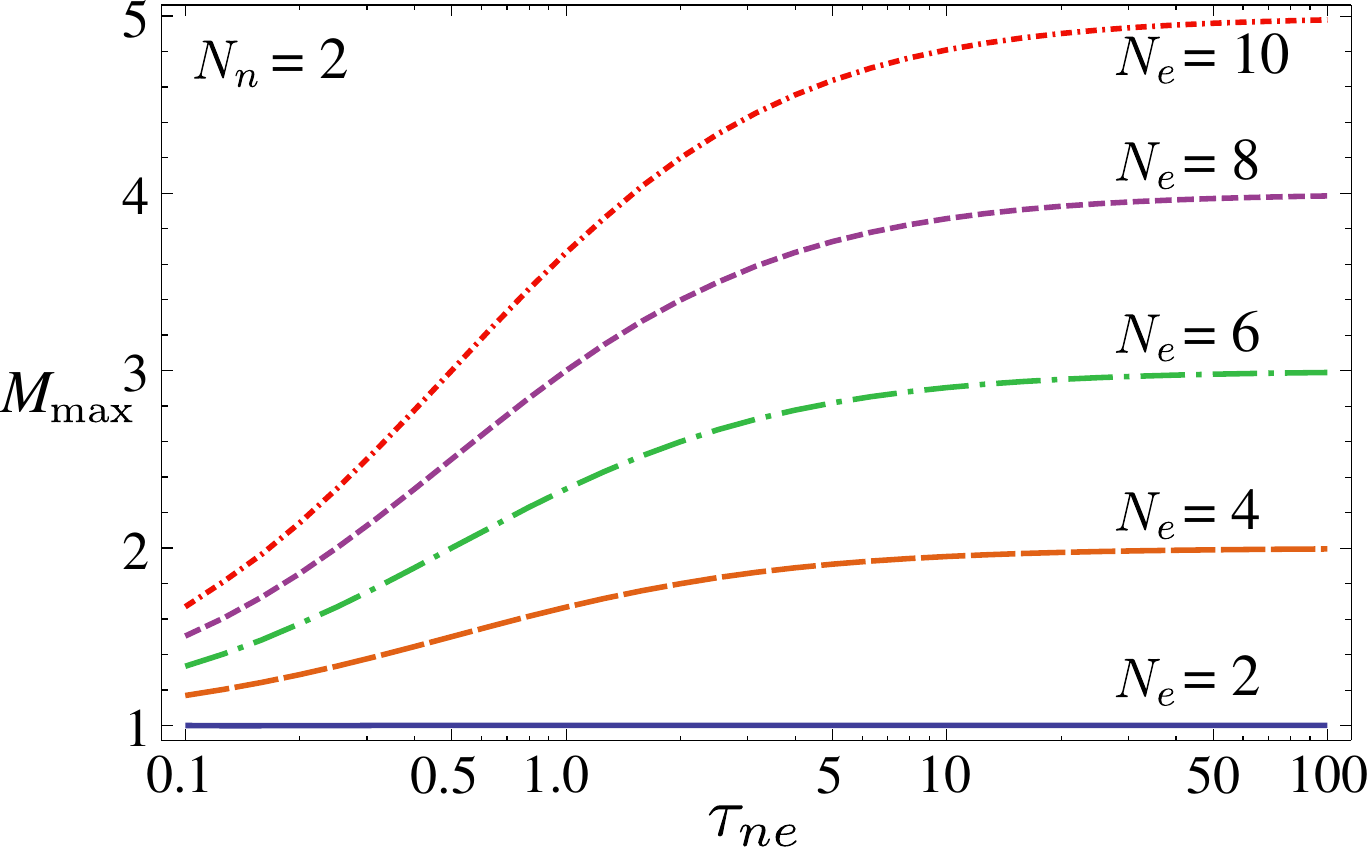}
\caption{(Color online) Mass ratio $M_\text{max}(\tau_{ne})$ which maximizes the nuclei-electrons entanglement as a function of the interaction $\tau_{ne}$, {\it i.e.} $M_\text{max}~~|~~\varepsilon(\tau_{ne},M)<\varepsilon(\tau_{ne},M_\text{max}), ~~ \forall M>0$. We plot $M_\text{max}(\tau_{ne})$ for systems with two nuclei ($N_n=2$) and different numbers of electrons ($N_e=2,4,6,8,10$).}
\label{MaxDeviationInTau}
\end{figure}

In Figure~\ref{MaxDeviationInTau}, we show the value of the mass ratio $M$ which maximizes the entanglement $\varepsilon$ for a given interaction $\tau_{ne}$. For small values of $\tau_{ne}$, the maximum is always located in the interval $1<M<N_e/N_n$. Increasing $\tau_{ne}$, the maximum entanglement moves up to the extreme value $M=N_e/N_n$ ({\it i.e.} the two subsystems have equal masses) which is reached in the limit $\tau_{ne}\to\infty$. Moreover, for $N_n=N_e=2$ (solid blue line in Fig. ~\ref{MaxDeviationInTau}), maximal entanglement is achieved when $M=1$ for any interaction $\tau_{ne}$.

We now explore the entanglement features considering the number of particles of the system.
The potential function appearing in the Hamiltonian \eqref{ManyParticleHamiltonian}
is a quadratic function of the complete set of
vector positions ${X_1}, \ldots {X_{N_n}},
{x_1}, \ldots {x_{N_e}}$. Note that
the independent particle frequencies corresponding to ${X_i}^2$ and
${x_j^2}$, which are 
$\Lambda_n/2 = (1/2) + (\tau_{ne}/2)N_e + (\tau_{nn}/2)(N_n -1)$ and
$\Lambda_e/2 = (1/2) + (\tau_{ne}/2)N_n + (\tau_{ee}/2)(N_e -1)$ respectively,
grow linearly with $N_n$ and $N_e$, while the
pre-factors corresponding to the cross interaction terms
like ${X_i} \cdot {X_j}$,
${X_i} \cdot {x_j}$, and
${x_i} \cdot {x_j}$, do not, (see Eq.~\eqref{AMatrix}). For large numbers of particles the leading part of the Hamiltonian \eqref{ManyParticleHamiltonian} is of the form $(\Lambda_n/2) \sum_{i=1}^{N_n} {X_i}^2 + (\Lambda_e/2) \sum_{j=1}^{N_e} {x_j}^2 $. This form of the potential function describes a set of $N_n + N_e$ independent harmonic oscillators.

\begin{figure}[ht]
\includegraphics[scale=0.42]{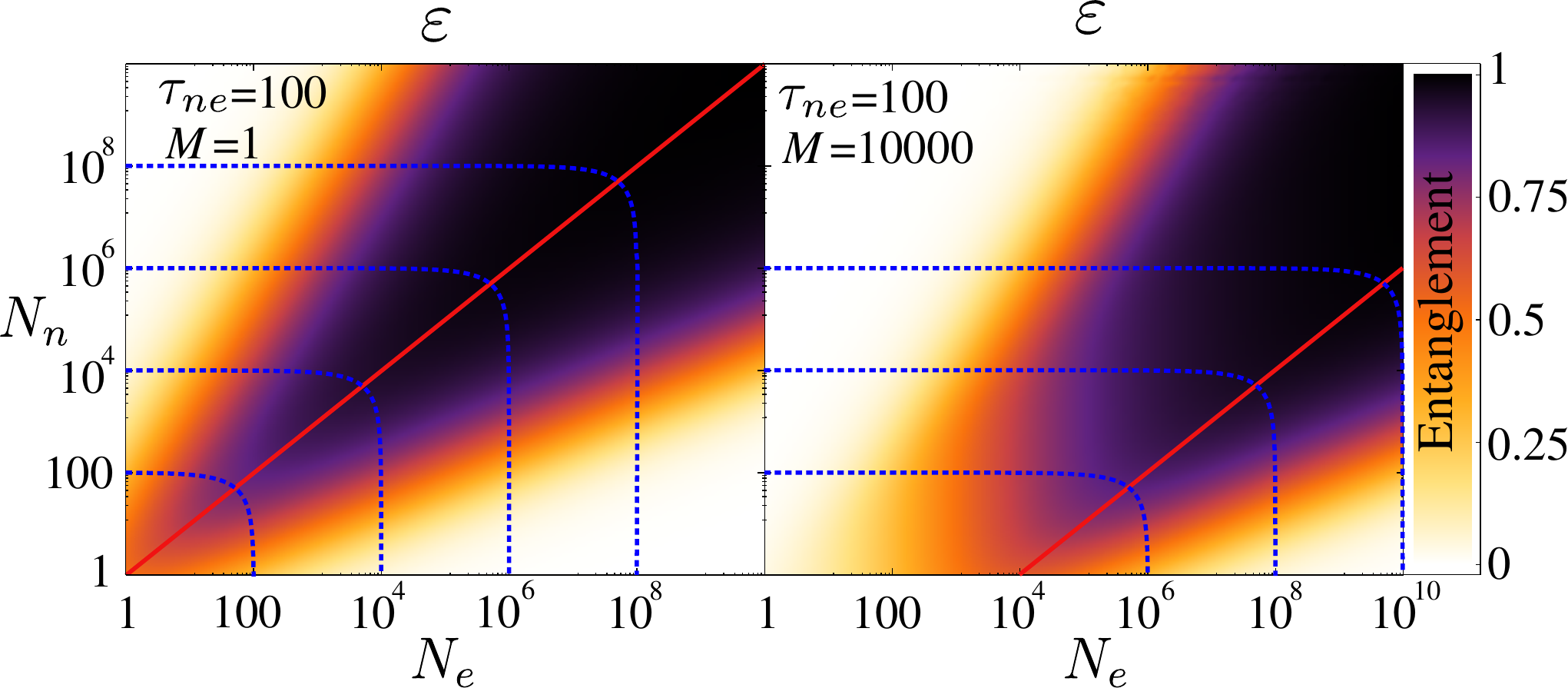}
\caption{(Color online) Nuclei-electrons entanglement, $\varepsilon(\tau_{ne},M,N_n,N_e)$ with $\tau_{ne}=100$, as a function of the number of nuclei $N_n$ and electrons $N_e$, for different interaction mass ratios. a) $M=1$, and b) $M=10000$. Solid red lines correspond to subsystems (nuclei and electrons) with equal masses, {\it i.e.} $M N_n = N_e$. Dotted blue lines correspond to systems with equal total mass $M_T= M N_n + N_e $.}
\label{EntNnNe}
\end{figure}

The number of correlation cross terms, $X_i\cdot x_j$, which contribute effectively to the reduced density matrix of the nuclei (or electrons), increases with $N_n \cdot N_e$. When $N_n \cdot N_e \simeq N_n + N_e$, the contributions of the cross terms are negligible. Therefore, the reduced density matrix can be approximated by the ground state associated with the independent many-particle potential, giving rise to non-entangled states. However, when $N_n \cdot N_e \gg N_n + N_e$, many cross terms induce correlations, in which case highly entangled states are obtained (see Fig.~\ref{EntNnNe}).

In the regime of large nucleus-electron interactions we can summarize that the region of higher entangled states is always located in the neighborhood of $N_n M = N_e$, (see red line in Fig.~\ref{EntNnNe}). Indeed, for systems with the same total mass $M_T= M N_n + N_e $ (dotted blue lines), the maximal entanglement is obtained when the subsystems have equal mass $M_T/2=N_n M = N_e$. If we gradually increase $N_n$ ($N_e$), while keeping $N_e$ ($N_n$) fixed, the entanglement fades away, but, increasing both $N_n$ and $N_e$ to infinity, the entanglement reaches its maximally possible value $\varepsilon=1$. This is illustrated in Fig.~\ref{EntNnNe} which is fully consistent with the above explanation. The particular case of $N$ particle with the same mass ($M=1$) was studied in Ref.~\cite{KoscikOkopinska2013} which shows that the maximal entanglement is obtained when the two subsystems have the same number of particles $N/2$ as depicted in
the blue lines of Fig.~\ref{EntNnNe} $a)$.

Finally, let us point out that the nuclei-electrons entanglement, $\varepsilon$, vanishes in the limit $M\to\infty$. This feature can be understood straightforwardly from the Born-Oppenheimer wavefunctions as we will show in section~\ref{Sec:BOManyParticleModel}.

\subsection{Single-particle entanglement: Nucleus and Electron entanglement}

\begin{figure}[ht]
\includegraphics[scale=0.42]{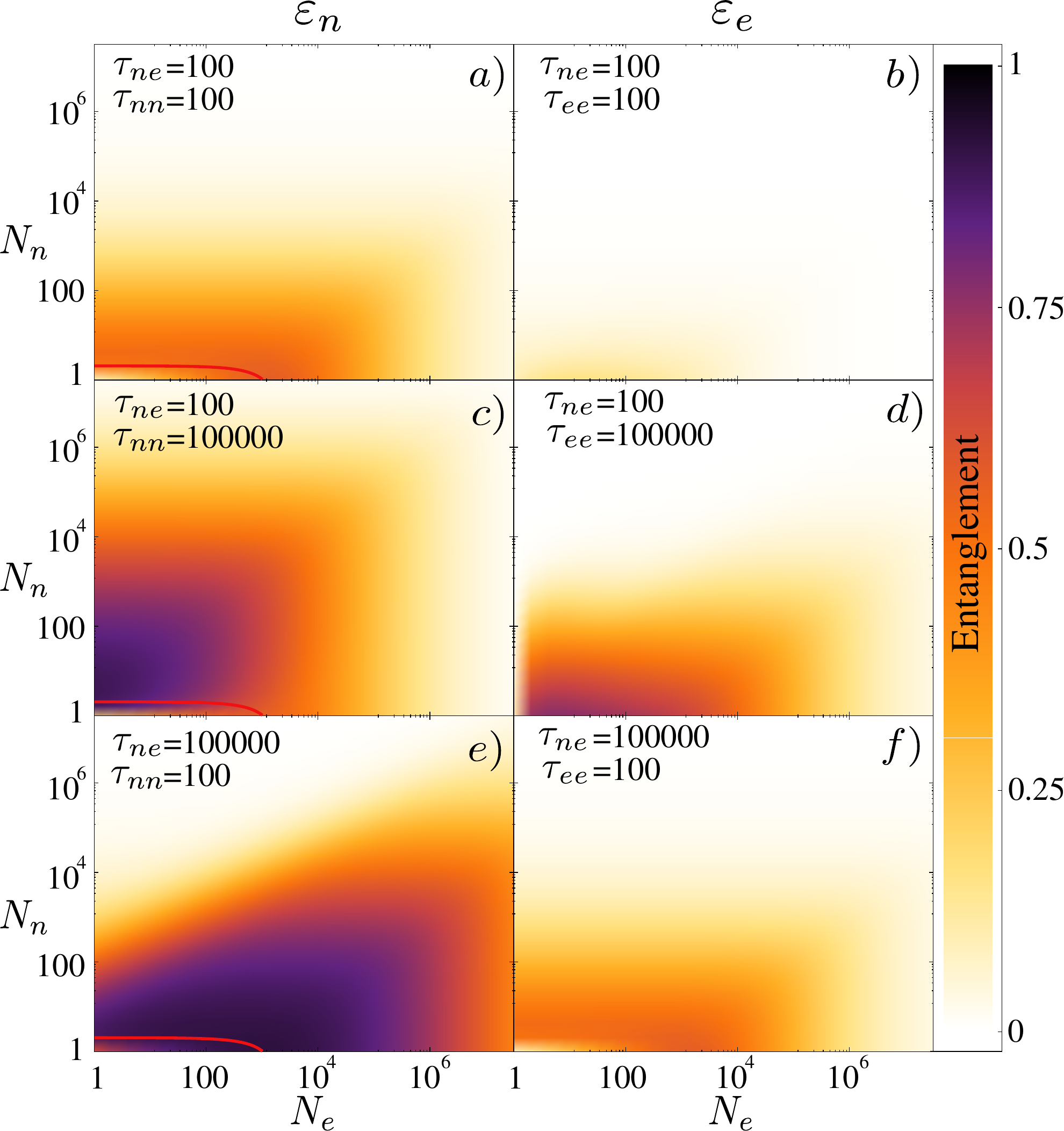}
\caption{(Color online) Nucleus entanglement, (left panels) $\varepsilon_n(\tau_{nn},\tau_{ne},M,N_n,N_e)$, and electron entanglement, (right panels) $\varepsilon_e(\tau_{ee},\tau_{ne},M,N_n,N_e)$, as a function of the number of nuclei $N_n$ and electrons $N_e$, electrons of the system for some relative interaction $\tau_{ne}$, $\tau_{ee}$, $\tau_{nn}$ and for the mass ratio $M=1000$. Solid red lines represents the cases of equal subsystem masses fulfilling the relationship $M = (N_n-1) M + N_e$. For the electron case there is no such relationship because we are implicitly assuming that $M > 1$.}
\label{EntNandE}
\end{figure}

Contrary to the nuclei-electrons entanglement increasing both $N_n$ and $N_e$ the single-particle entanglement vanishes, as shown in Fig.~\ref{EntNandE}. In this case, the number of cross terms of the Hamiltonian containing the correlations which effectively contribute to the reduced density matrix of a single particle is $N_n+N_e-1$. Increasing $N_n$ or $N_e$, the contribution of the cross terms becomes negligible and the single-particle reduced density matrix corresponding to a nucleus (electron) approaches the projector on the ground state associated with the single-particle potential $(\Lambda_n/2){X}^2$ ($(\Lambda_e/2){x}^2$). The reduced single-particle density matrices of a nucleus or an electron approaches to a pure states disentangled from the rest of the system.

When all particles interact with the same strength, $\tau_{ne}=\tau_{nn}=\tau_{ee}=\tau \ge 1$, so that we do not privilege any interaction, a nucleus is always more correlated with the rest of the system than an electron, {\it i.e.} for all $N_n$, $N_e>1$, a single particle entanglement fulfils
\eq
\varepsilon_n  > \varepsilon_e & \text{if~~} M>1 \\
\varepsilon_n  = \varepsilon_e = \varepsilon_{1} & \text{if~~} M=1
\en
where
\eq
\varepsilon_{1} = 1-\frac{\left(1+\sqrt{A}\right) A^{1/4}}{\sqrt{\left(\sqrt{A}+A-\tau \right) \left(1+\sqrt{A}+\tau \right)}}
\en
and $A= 1 + N_e \tau + N_n \tau$. The hierarchy on the entanglement reveals the composite nature of the particles \cite{TichyBouvrie2012a}. Here, heavier elementary particles of a composite particle (or molecule) are more entangled, ``in a hard core,'' than the light ones, which are more likely to exhibit the composite nature of the ``molecule''.

The electron entanglement is always a decreasing function of $M$, $\varepsilon_e(M)\le \varepsilon_e(1)$, which highlights the confining effect of nuclei on the electrons (entanglement decreases with the confinement \cite{BouvrieMajteyEtal2012}). On the other hand, if the system has more nuclei than electrons, $N_n\gg N_e$, and if $\tau_{ne} \le \tau_{nn}$, then the nucleus entanglement $\varepsilon_n$ is independent of $N_e$, $\tau_{ne}$ and $M$, {\it i.e.} nuclei do not feel electrons and $\varepsilon_n$ is that given by a system of $N_n$ particles with the same mass.

When all interactions between particles are equal, $\tau_{ne}=\tau_{ee}=\tau_{nn}=100$, we can observe in Fig.~\ref{EntNandE}$a$ and \ref{EntNandE}$b$ that $\varepsilon_n > \varepsilon_e$. Similar trends are found for comparable interactions in Figs.~\ref{EntNandE}$b$ and \ref{EntNandE}$c$, and Figs.~\ref{EntNandE}$e$ and \ref{EntNandE}$f$. In general, we observe that the entanglement decreases with the number of particles $N_n$ and $N_e$ but, in Fig.~\ref{EntNandE}$e$, we found remarkably different behavior from the entanglement: for a fixed $N_n$, the entanglement displays a maximum when $N_e$ increases. In general, a nucleus is more entangled with the other nuclei, but due to the large interaction with the electrons the entanglement increases with $N_e$ (becuase there are more electrons highly correlated with the nucleus) until the cross terms are negligible as compared to the particle-independent terms.

\section{Born-Oppenheimer approximation and nuclei-electrons entanglement}
\label{Sec:BOManyParticleModel}

The study of entanglement as a function of the masses of the constituent particles of composite systems
naturally leads to considerations of the connection between entanglement
and the celebrated Born-Oppenheimer (BO) approximation.
The validity of the BO approximation \cite{BornOppenheimer1927} is closely related to the masses of the particles and probably constitutes the most fundamental approximation in quantum chemistry \cite{McQuarrie2007} and in molecular physics \cite{Demtroder2007}. From a practical point of view, the BO approximation allows us to compute the electronic structure of a molecule for a given configuration of its nuclear part.

The physical motivation behind the BO approximation is that the nuclei are much heavier than the
electrons. Therefore, one can consider the nuclei position coordinates $\mathbf{X}$ as parameters that define the effective Hamiltonian
for the electrons. For any fixed configuration of the nuclei, one
has to solve a Schr\"{o}dinger equation that involves only the
electronic degrees of freedom. The eigenvalues and eigenfunctions
depend on the particular nuclear configuration. Once one has
solved the electronic Schr\"odinger equation, the effective
Hamiltonian for the nuclei can be obtained by adding
the electronic eigenenergy to the nuclear Schr\"odinger equation.

\subsection{BO Many-particle wavefunctions}

BO approximation, assumes that the heavy particles (nuclei) move more slowly than the light ones (electrons), and it is therefore common to use the electronic stationary Born-Huang expansion \cite{BornHuang1954,YoneharaHanasakiEtal2012}
\eq
\Psi(\mathbf{X},\mathbf{x},t)=\sum_n F_n(\mathbf{X},t)\phi_n(\mathbf{X},\mathbf{x}),
\en
for either adiabatic and diabatic theories. Here, we focus on the zeroth adiabatic approximation with a time-independent potential function, and hence the wavefunction reduces to the BO Ansatz \cite{BornOppenheimer1927}
\eq
\label{BOWaveFunction}
\Psi^{BO}_{\mathbf{s},\mathbf{q}}(\mathbf{X},\mathbf{x})= F_\mathbf{s}(\mathbf{X})\phi_\mathbf{q}(\mathbf{X},\mathbf{x}),
\en
where $\mathbf{s}$ and $\mathbf{q}$ denotes the quantum states of the nuclei and electrons, respectively. In this approximation, the electrons move adiabaticaly in the field of fixed nuclei at the positions $\{X_i\}$.

In order to obtain more detailed insight, let us apply the approximation to a, ``molecule,'' composed of $N_n$ nuclei and $N_e$ electrons, considered here. The time-independent Schr\"odinger equation for the system \eqref{ManyParticleHamiltonian} is
\eq
\label{hamiBO}
[T_n+T_e+V]\psi(\mathbf{X};\mathbf{x}) = E\psi(\mathbf{X};\mathbf{x}),
\en
where $T_n$ and $T_e$ denote the kinetic energy operator for the nuclei and electrons, respectively, and $V$ the total potential energy of the system.

Using the Ansatz \eqref{BOWaveFunction}, the electronic wave equation is given by
\eq
\label{ElecHamiBO} (T_e+V)\phi_\mathbf{q}(\mathbf{X};\mathbf{x}) =
E^{elec}_\mathbf{q}(\mathbf{X})\phi_\mathbf{q}(\mathbf{X};\mathbf{x}),
\en
where $E^{elec}_\mathbf{q}$ and the wavefunction $\phi_\mathbf{q}$ for each electronic state $\mathbf{q}$
depends parametrically on the nuclear coordinate $\mathbf{X}$. The nuclear wavefunction $F_\mathbf{s}(\mathbf{X})$ satisfies
\begin{equation}
\label{NucHamiBO}
 \left[T_n+E^{elec}_\mathbf{q}(\mathbf{X})-E\right]F_\mathbf{s}(\mathbf{X})=0.
\end{equation}

Again the solution is found in terms of Jacobi coordinates \eqref{JacobiChangeRelative2}, \eqref{JacobiChangeRelative1}, and \eqref{JacobiChangeCentermass}. The nuclear eigenfunctions are given by
\eq
F_\mathbf{s}(\mathbf{X}) = \Phi^{\beta^{(n)}_{N_n}}_{s_{N_n}}\left(\sqrt{N_n} R_{N_n} \right)\prod_{j=1}^{N_n-1} \Phi^{\beta^{(n)}}_{s_j}\left(R_j\right)
\en
and the electronic eigenfunctions by
\eq
\label{ElecWF}
\phi_\mathbf{q}(\mathbf{X};\mathbf{x}) = \Phi^{\beta^{(e)}_{N_e}}_{q_{N_e}}\left(\sqrt{N_e} r_{N_e} - \delta \sqrt{N_n} R_{N_n}\right) \times \nonumber \\ \times \prod_{i=1}^{N_e-1} \Phi^{\beta^{(e)}}_{q_i}\left(r_i \right),
\en
where the function $\Phi_{\nu}^{\beta(j)}(y)$ is given by Eq. \eqref{harmonic-solution}, the frequencies $\beta^{(n)}$ and $\beta^{(e)}$ are given by Eqs.~\eqref{betan} and \eqref{betae} respectively, and with $\delta = \frac{\tau_{ne}\sqrt{N_nN_e}}{\sqrt{M}(1+N_n\tau_{ne})}$. The frequencies corresponding to the center of mass coordinates are given by
\eq
\beta_{N_n}^{(n)} = \frac{1+(N_n+N_e)\tau_{ne}}{M(1+N_n \tau_{ne})}, \ \beta_{N_e}^{(e)} = 1+N_n\tau_{ne}.
\en

To test the validity of the BO approximation, we use the overlap between the exact \eqref{1DExactWaveFunctionState} and approximate \eqref{BOWaveFunction} wavefunctions given by
$\Theta_{u_1,u_2,\mathbf{n},\mathbf{e}|\mathbf{s};\mathbf{q}} = \langle u_1,u_2,\mathbf{n},\mathbf{e}|\mathbf{s};\mathbf{q}\rangle$,
where we have denoted the state associated to the wavefunction \eqref{BOWaveFunction} by $|\mathbf{s};\mathbf{q}\rangle=|s_1,...,s_{N_n};q_1,...,q_{N_e}\rangle$ such that
\eq
\label{1DBOWaveFunction}
\Psi^{BO}_{\mathbf{s},\mathbf{q}}(\mathbf{X},\mathbf{x}) = \langle \mathbf{X},\mathbf{x} |\mathbf{s};\mathbf{q}\rangle.
\en
For the ground state, $\Theta_{gs}=\Theta_{0,0,0,0|0;0}$, one has to evaluate the integral
\eq
\label{overlap1D}
\Theta_{gs} =
\int_{-\infty}^{\infty} dR_{N_n}dr_{N_e} ~ \prod_{l=1}^2 \Phi_{0}^{\beta_l}\left(U_l(R_{N_n},r_{N_e})\right)  \times \nonumber  \\
\times \Phi^{\beta^n_{N_n}}_{0}(\sqrt{N_n}R_{N_n}) \Phi^{\beta^e_{N_e}}_{0}(\sqrt{N_e} r_{N_e} - \delta \sqrt{N_n} R_{N_n}) .
\en

\subsection{Wavefunctions and Entanglement}

The eigenfunctions corresponding to the relative coordinates, Eqs.~\eqref{JacobiChangeRelative2} and \eqref{JacobiChangeRelative1}, in the BO wavefunction \eqref{BOWaveFunction} are exactly the same as in the exact solution \eqref{ExactWaveFunctionIndepOsci}. Indeed, all correlations between nuclei and electrons are again induced by their respective centers of mass, but in a different way; now they are all embedded in the electronic wavefunction \eqref{ElecWF}.

The harmonic frequencies associated with the nuclei, $\sqrt{\beta^{(n)}}$ and $\sqrt{\beta^{(n)}_{N_n}}$, decrease with $\sqrt{M}$, which fits the contribution of the counterpart (nuclei and electron) to the wavefunction, but not to the electronic one, $\sqrt{\beta^{(e)}}$ and $\sqrt{\beta^{(e)}_{N_e}}$. At first reading, we can infer that the approximation becomes increasingly accurate as $M$ increases. But, as shown later, the number of particles also strongly affects validity of the approximation.

We have seen in the previous sections \ref{Sec:3PaExactModel} and \ref{Sec:ManyPartNucElecEnt} that the nuclei-electrons entanglement $\varepsilon$ vanishes in the limit $M\to\infty$, but also taht the accuracy of the BO approximation is maximized.
Increasing $M$, the contribution of the nuclei to the total wavefunction becomes more relevant than the electronic one. In the limit $M\rightarrow\infty$, the nuclear wavefunction $F_{\mathbf{s}}(\mathbf{X})$ is a delta-like function at the nuclear
positions. Therefore, electrons ``feel'' nuclei as an external confining potential, and nuclei are virtually unaffected by electrons, thus denying any possibility of entanglement between nuclei and electrons.

By their very definition, the wavefunctions of states without entanglement between nuclei and electrons ($\varepsilon=0$) factorize as a product of nuclear and electronic wavefunctions, $\Psi(\mathbf{X},\mathbf{x})=F(\mathbf{X})\phi(\mathbf{x})$. This wavefunction is sometimes referred to as a completely adiabatic state. In such case, the assumption \eqref{BOWaveFunction} is perfectly fulfilled, and the BO approximation provides the exact wavefunction which coincides with the completely adiabatic state. Therefore, the entanglement can be used to assess the validity of the BO approximation as far as non-entangled states imply maximal accuracy of the approximation.

Besides the inherent entanglement in electronic adiabatic states \eqref{BOWaveFunction}, in nonadiabatic approximations there is another source of entanglement due to the bifurcation of the wavefunction \cite{YoneharaHanasakiEtal2012}. The nonadiabatic time evolution is governed mathematicaly by the state bifurcantion
\eq
F_1(\mathbf{X},t_b)\phi_1(\mathbf{X},\mathbf{x}) \rightarrow & F_1(\mathbf{X},t_a)\phi_1(\mathbf{X},\mathbf{x}) + ~~~~~~~\\ \nonumber & + F_2(\mathbf{X},t_a)\phi_2(\mathbf{X},\mathbf{x}) + \cdots
\en
Contrary to the zeroth adiabatic states for which the entanglement is time-independent even for time-dependent potential functions, the entanglement of nonadiabatic electronic states is, in general, time-dependent. This last entanglement could be cumbersome to compute analytically and it is beyond the scope of the current work.

\subsection{The $H_2^+$-like ``molecule"}
\label{Sec:BOH2+}

The simplest molecular case is the ground state of the $H_2^+$-like molecule ($N_n=2$ and $N_e=1$). For such a system, we compute and analyze the overlap measure \eqref{overlap1D}, as well as the nuclei-electrons entanglement $\varepsilon$, with both exact and approximate methods.

\begin{figure}[h]
\includegraphics[scale=0.55]{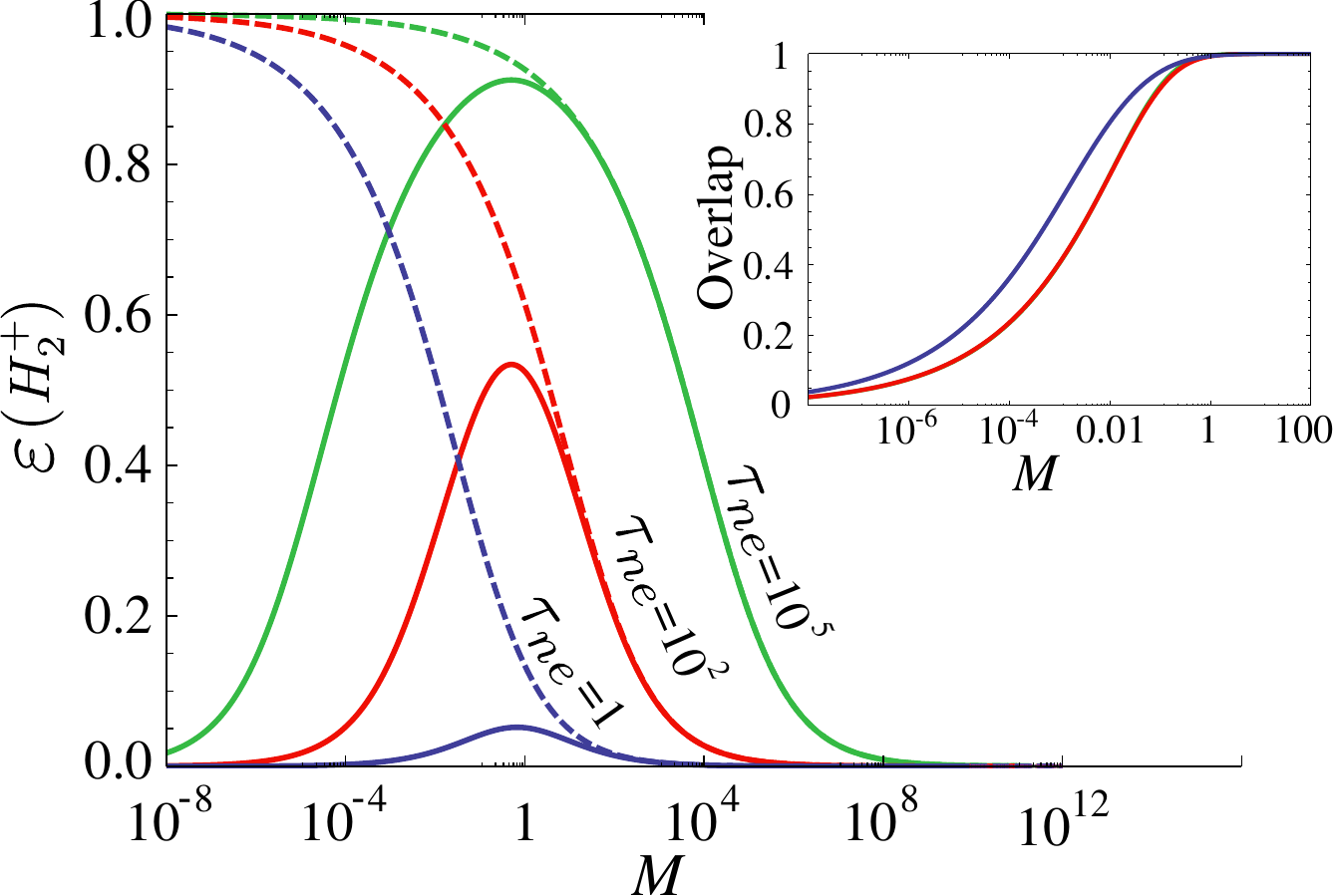}
\caption{(Color online) Nucleus-Electrons ground state entanglement, $\varepsilon(\tau_{ne},M,N_n,N_e)$ with $N_n=1$ and $N_e=2$ ($H_2^+$), exactly computed (solid lines) and with the Born-Oppenheimer approximation (dashed lines) as a function of the mass ratio $M$. In the inset figure we plot the ground state overlap of the ground state, $\Theta_{gs}(H_2^+,\tau_{ne},M)$ Eq.~\eqref{overlap1D}, as a function of $M$. 
}
\label{Helium-likeAtomBO}
\end{figure}

For a large interaction $\tau_{ne}\gg1$, the BO approximation can accurately describe high entanglement states over a wide range of masses, $M\gg1$. The approximated entanglement becomes more accurate for increasing values of $M$ until it vanishes in the limit $M\to\infty$, (see Fig.~\ref{Helium-likeAtomBO}).

Then we give the interaction and mass orders roughly, to get closer to a more realistic $H_2^+$-like molecule viewpoint. The molecular size is of the order of the Bohr radius $a_0$, and the usual atomic trap size of $b\sim 10^6a_0$, So that the relative interaction strength is of the order of $\tau_{ne}=\lambda_{ne}/k=(b/a_0)^4\sim 10^{24}$. Taking into account this, and the fact that the proton-electron mass ratio is of the order $M_{pe}\approx2000$, one finds that the ground state of a $H_2^+$-like molecule in a commonly harmonic trap is highly (nuclei-electrons) entangled . The trace of the square marginal density matrix is of the order of $\text{Tr}[\rho_n^2] \approx 8.8 \cdot 10^{-6}$. In such a case, the BO turns out to be a good approximation, for which we obtain similar results of entanglement with a relative error of $0.014\%$.

More realistic wavefunctions of molecules require very intricate numerical calculations which add to the computational expenses of the linear entropy. In addition to the BO approximation, one may employ another commonly used approach (see Ref.~\cite{ChudzickiOkeEtal2010}) to compute the linear entropy. Considering the integrals in \eqref{deftraza} and the BO Ansatz \eqref{BOWaveFunction}, one can assume that the nuclear wavefunction $F(\mathbf{X'})$ can be approximated by $F(\mathbf{X})$ when $|\mathbf{X}-\mathbf{X'}|$ is of the order of the atom size, whenever the relative nuclei-electrons interaction strength $\tau_{ne}$ is large enough and $M$ is not very large ($1\ll M\ll \tau_{ne}$).

This last approximation reduces the dimensions of the integrals involved in the linear entropy. The use of both approximation could significantly simplify both numerical and analytical computations of the entanglement amount in more complex and realistic systems,  e.g., for the $H_2^+$ molecule model used here we obtain
\eq
\varepsilon(H_2^+) \approx 1-\frac{\sqrt{2+4 \tau_{ne}} (M+3 M \tau_{ne})^{1/4}}{2 \tau_{ne}}.
\en

\begin{figure}[h]
\includegraphics[scale=0.55]{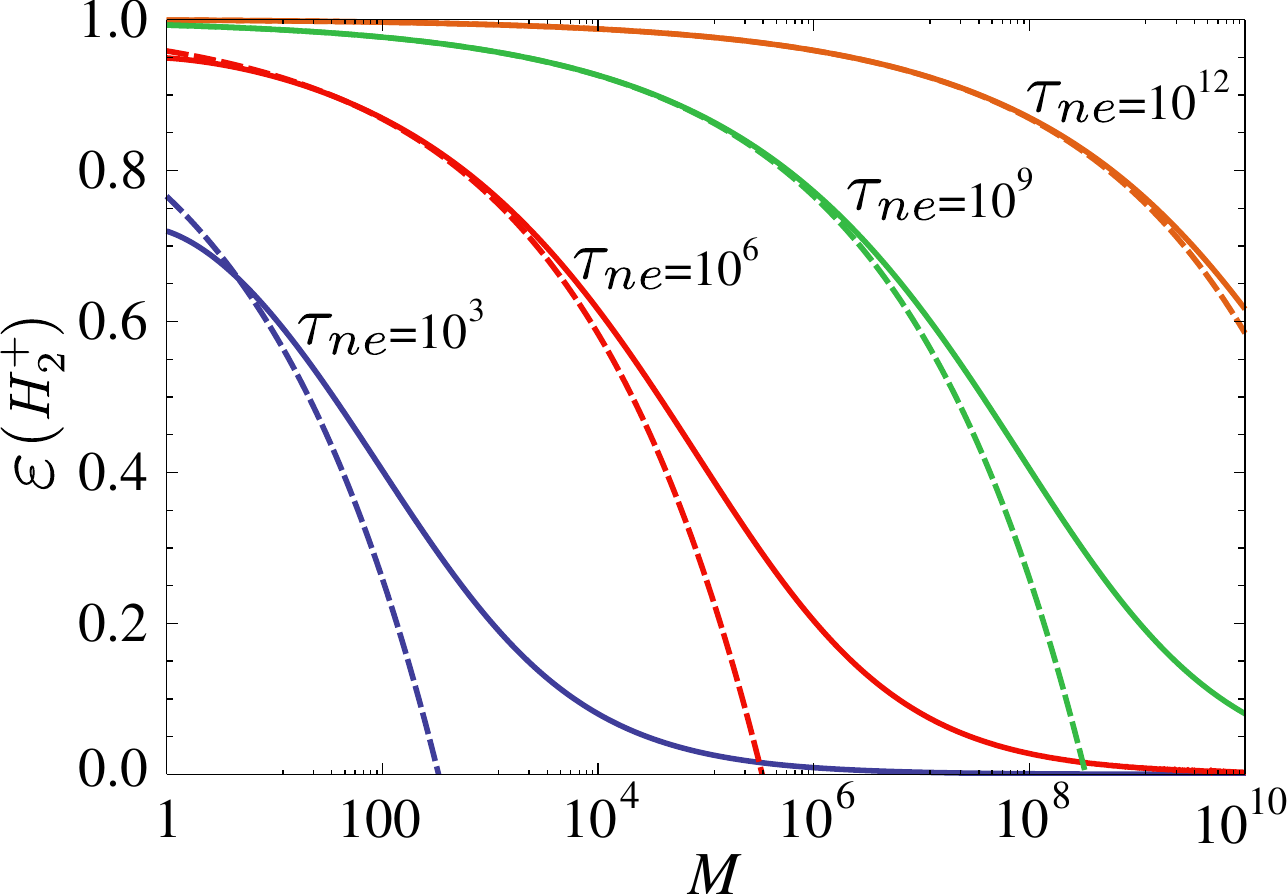}
\caption{(Color online) Dependence of the nucleus-electrons entanglement, $\varepsilon(H_2^+,\tau_{ne},M)$, on the mass ratio $M$ for several values ​​of $\tau_{ne}$. We compare the entanglement computed exactly (solid lines) with that computed by means of the Born-Oppenheimer approximation, using the assumption that $|\mathbf{R}-\mathbf{R'}|$ is of the order of the atom size (dashed lines).}
\label{Helium-likeAtomBOandRApproach}
\end{figure}

Moreover, we cannot lose sight of the short mass range which is valid for both approximations; for very large $M$ this last approximation in \cite{ChudzickiOkeEtal2010} gets worse, and for $M\sim 1$ the BO approximation fails (see Fig. \ref{Helium-likeAtomBOandRApproach}).
Thanks to the electron-proton mass ratio and the physical range of the electron-proton relative interaction in a common harmonic trap, these two approximations can be used together to compute the nuclei-electrons entanglement of atoms and molecules. Indeed, in this mass and interaction ranges of the $H_2^+$ ground state the relative error of the linear entropy is almost the same ($0.014\%$), which is mainly due to the BO approximation itself.

\subsection{Many-particle Systems}
\label{SubSec:NParticlesBO}

As noted in previous sections \ref{Sec:BOManyParticleModel} and \ref{Sec:BOH2+} the validity of the BO approach is closely related to the mass ratio of the particles (see Fig. \ref{Helium-likeAtomBO}), however, we found also a strong dependence on the number of particles composing the system. The accuracy of the approximation increases with the total mass ratio between the full subsystems of nuclei and electrons, {\it i.e.}, when $\gamma=\frac{MN_n}{N_e}$ increases.

\begin{figure}[h]
\includegraphics[scale=0.45]{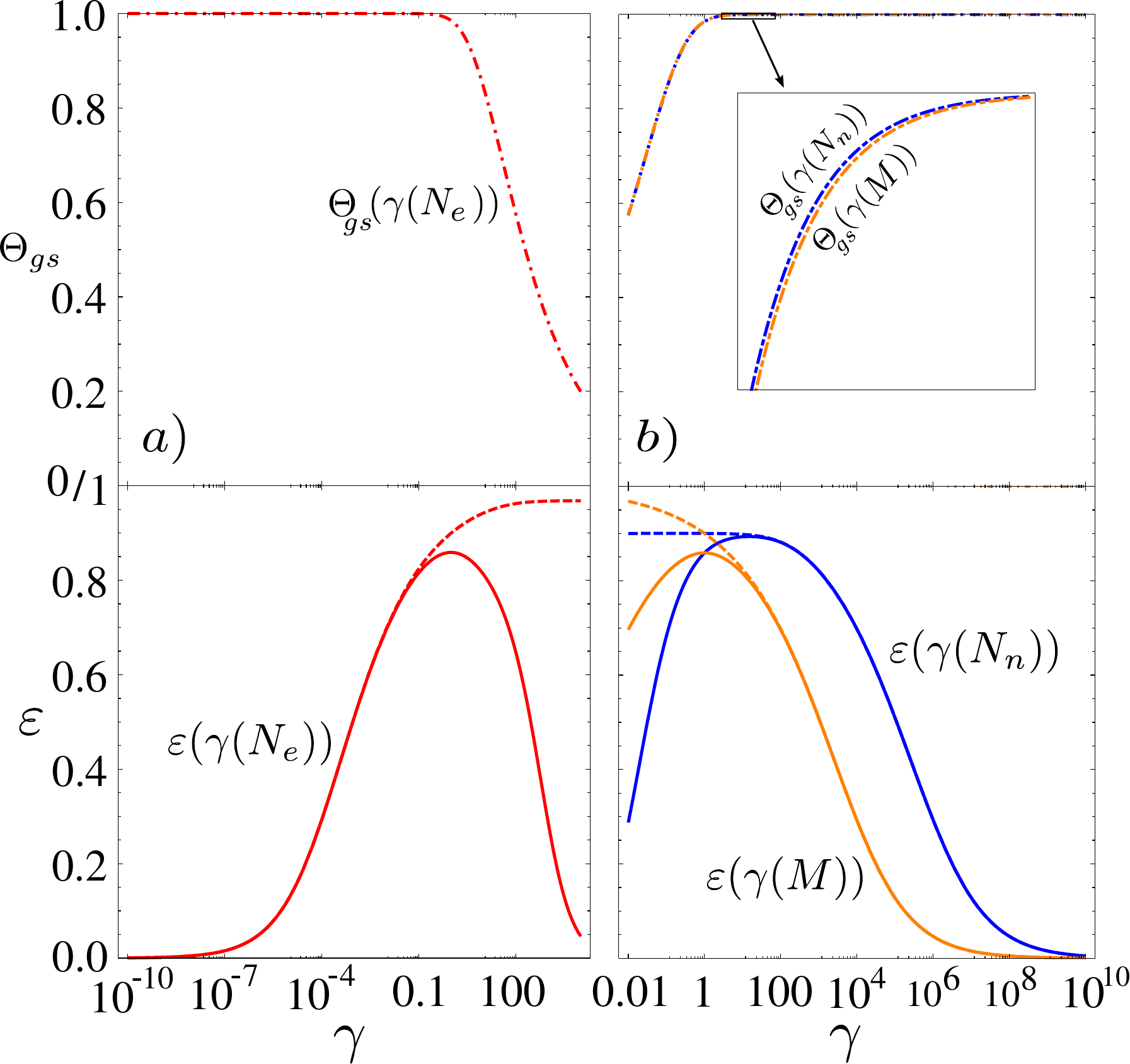}
\caption{(Color online) Comparison of the exact (solid lines) and BO approximation (dashed lines) ground state entanglement, $\varepsilon(\tau_{ne},M,N_n,N_e)$ (lower panels), and their wavefunctions overlap (dotted dashed lines in the upper panels) as a function of $\gamma=\frac{MN_n}{N_e}$; $a)$ fixed $N_n=100$ and $M=100$, $b)$ fixed $N_e=10000$ and $M=100$ (blue lines), and $N_e=10000$ and $N_n=100$ (orange lines).}
\label{ManyParticlesBO}
\end{figure}

Fixing any two parameters (of $N_n$, $N_e$ and $M$), the ground state overlap is maximal, $\Theta_{gs} \approx 1$, when $\gamma\gg1$, (see upper panel in Fig.~\ref{ManyParticlesBO}$b)$. This can be inferred from the exact \eqref{1DExactWaveFunction} and BO \eqref{BOWaveFunction} wavefunctions. On the other hand, when $\gamma<1$, one can deal with the complementary BO approximation and solve the nuclei differential equation by considering electron position coordinates as fixed parameters. In such a case, the accuracy of the BO approximation increases when $\gamma$ decreases and the overlap approach to unity, $\Theta_{gs} \approx 1$ for $\gamma\ll1$, (see upper panel in Fig.~\ref{ManyParticlesBO}$a)$. We can therefore conclude that the decisive parameter for the accuracy of the BO approximation is the total mass ratio $\gamma$, and not just the mass of the constituent
particles. Indeed, all curves of the overlap as a function of $\gamma$ collapse to the same in the limit $\tau_{ne}\to\infty$.

The nuclei-electron entanglement vanishes in the limits $\gamma \to 0,\infty$ and is maximal when $\gamma\simeq 1$. This fact is fully consistent with the kinematic intuition mentioned above; when the interaction between the compounds of the two subsystems is not negligible, the entanglement is greater for subsystems with similar masses and decreases as the masses become more different (see lower panels in Figs.~\ref{ManyParticlesBO}$a)$ and $b)$). The subsystems are more prone to influence each other when their total masses are similar.

Both, the validity of the BO approximation and nuclei-electrons entanglement are therefore governed by the mass ratio of the full subsystems $\gamma$. The BO approximation is able to accurately describe highly entangled states in a wide range of $\gamma\gg1$, {\it e.g.} for atoms and molecules: however, the completely adiabatic approximation does not provide a correct description of the states and hence of its associated entanglement. In the limiting cases of $\gamma \to 0,\infty$ one always has non-entangled states and maximal accuracy. In this limit, the completely adiabatic state coincides with the BO and the exactly computing state. Therefore, beyond the wavefunction overlap, entanglement can be used to assess the validity of the BO approximation and to discern whether or not adiabatic theory is completely applicable. In other words, non-entangled states legitimize the use of BO as well as the completely adiabatic approximations to compute the wavefunction of the system . This validity test based on the 
entanglement becomes stronger than the overlap as the interaction increases and it is inefficient for small interactions $\tau_{ne} \lesssim 1$.

\section{Conclusions}
\label{Sec:Conclusions}

In this paper we investigated the entanglement
properties of an one-dimensional $N$-particle system consisting of $N_n$ ``nuclei'' and
$N_e$ ``electrons'' which interact harmonically with each other. Moreover, they are confined by an
harmonic external potential. 

As a general trend, we found that entanglement increases with the
interaction between particles, approaching its maximal
possible value in the limit of an infinitely large interaction.
Excited states have been studied in the three-particle case, which exhibits a finite amount of entanglement
in the limit of vanishing interactions due to the degeneracy of the energy levels of the Hamiltonian
describing non-interacting particles.

In the many-particle case we have only investigated the ground state. For sufficiently large $N_n$ or $N_e$ values, the entanglements $\varepsilon$ , $\varepsilon_n$ and $\varepsilon_e$ vanish, but, by increasing simultaneously the number of both ``nuclei'' and ``electrons,'' $\varepsilon$ tends to its maximal value while $\varepsilon_n$ and $\varepsilon_e$ vanish. We have shown that this is due to the number of correlated cross terms of the Hamiltonian contributing to the reduced density matrix.
When the ``nuclei-electrons'' interaction is large ($\tau_{ne}\gtrsim100$), the ``nuclei-electrons'' entanglement can be understood kinematically, {\it i.e.} $\varepsilon$ displays a maximum when the masses of
the two considered subsystems are similar and vanishes when the subsystems have very different masses.

In summary, when it comes to mass and entanglement, interacting
parts of the model studied here exhibit a ``like-for-like"
behavior: when the system is partitioned into two
interacting subsystems, these parts tend to be highly entangled
with each other when they have similar masses. It would be
interesting to investigate to what extent this is a universal
trend verified by composite quantum systems.

We explored the connections mass-entanglement by means of the Born-Oppenheimer approximation. This approximation makes evident that by increasing the particle mass ratio $M$, the nuclear density approaches a Dirac delta function located at the ``nuclei'' positions and thus loses any possibility of entanglement. To estimate the quality of this approximation, we
studied the overlap between the exact wavefunctions and the BO
ones. Regarding the size of the system, both the overlap and the ``nuclei-electrons'' entanglement are governed by the total mass ratio of the subsystems $\gamma$; in the limits $\gamma\to 0,\infty$ the entanglement vanishes and the overlap is maximum. This allow us therefore, to assess the validity of the BO approximation via entanglement.

This result is fully consistent with what happens in
quantum chemistry and molecular physics, where the mass of ``nuclei'' is
 indeed much larger than the ``electron'' mass. In this regime, the
 Born-Oppenheimer approximation applies. It is worth stressing,
 however, that the Born-Oppenheimer Ansatz does not constitute a
 zero-entanglement approximation. In fact, we have shown in
 the present work that the Born-Oppenheimer approximation
 provides a good description of the system even in cases
 where it exhibits an appreciable amount of entanglement.
 Then, we can conclude that entanglement is not allways related to complexity. The Born-Oppenheimer approximation is computationally very efficient and we showed that, in spite of its simplicity,
 it describes certain entanglement features
%  and despite this fact, we showed that it describes certain entanglement features.

\acknowledgments{The authors gratefully acknowledge the MINECO grant FIS2011-24540 and the excellence grant FQM-7276 and FQM-207 of the Junta de Andaluc\'ia. A.P.M acknowledges support by GENIL through the YTR-GENIL Program and by CAPES/CNPQ through the BJT Ci\^encia sem Fronteiras Program. M.C.T. gratefully acknowledges support by the Alexander von Humboldt--Foundation through a Feodor Lynen Fellowship.}

\appendix

\begin{appendix}
\section{Derivation of the exact eigensolutions}
\label{Sec:AppendixA}

In this appendix we derive in detail the exact eigenfunctions and eigenenergies of the Hamiltonian \eqref{ManyParticleHamiltonian}.
To solve analytically the Schr\"odinger equation $H_x|\Psi\rangle=E_x|\Psi\rangle$ we introduce the dilatation coordinate change for nuclei
\eq
\label{dilatation}
X_j \rightarrow \frac{X'_j}{\sqrt{M}},
\en
which allows us to express the one-dimensional Hamiltonian, $H'_x=H_x,$ in terms of an $N\times N$ interaction matrix $A$
\eq
\label{ManyParticleHamiRescaled}
H'_x=\sum _{j=1}^{N_n} \frac{{P'_j}^2}{2}+\sum _{i=1}^{N_e} \frac{{p_i}^2}{2} + \frac{1}{2}\sum _{j=1}^{N_n}A_{j,j} {X'_j}^2 + \\
\frac{1}{2}\sum _{i=1}^{N_e} A_{N_n+i,N_n+i} {x_i}^2 - \sum _{j=1}^{N_n} \sum _{i=1}^{N_e} A_{j,N_n+i} {X'_j}{x_i} - \nonumber \\
\sum _{i=1}^{N_e} \sum _{j=i+1}^{N_e} A_{N_n+i,N_n+j}{x_i}{x_j} - \sum _{i=1}^{N_n} \sum _{j=i+1}^{N_n} A_{i,j} {X'_i}{X'_j}. \nonumber
\en
The elements of $A$ are given by
\eq
\label{AMatrix}
A_{j,j} &=& \frac{(1+N_e\tau_{ne}+(N_n-1) \tau_{nn})}{M},\nonumber \\
A_{i+N_n,i+N_n} &=& (1+N_n \tau_{ne}+(N_e-1)\tau_{ee}),\nonumber \\
A_{j,i+N_n} &=& - \frac{\tau_{ne}}{\sqrt{M}} ,\nonumber \\
A_{j,l} &=& - \frac{\tau_{nn}}{M} \text{ \ \ for} \ \ j \neq l,\nonumber \\
A_{i+N_n,m+N_n} &=& - \tau_{ee} \text{ \ \ for} \ \ i \neq m,
\en
where the indices $j$ and $l$ ($i$ and $m$) run between 1 and $N_n$ (1 and $N_e$) and refer to nuclear (electronic) coordinates.

The coordinates that allow us to rewrite the system Hamiltonian \eqref{ManyParticleHamiRescaled} in a fully separable form are given by the eigenvectors of the interaction matrix $A$. It has $N-2$ degenerate values. For the corresponding eigenvector we choose the Jacobi variables for electrons $\{ r_1, \ldots, r_{N_e -1}\}$ and nuclei $\{R_1, \ldots, R_{N_n - 1}\}$
\eq
\label{JacobiChangeRelative2}
R_j(X'_1,...,X'_{j+1})&=&\sum_{k=1}^{j} \frac{X'_k - X'_{j+1}}{\sqrt{j+j^2}} ,\\
\label{JacobiChangeRelative1}
r_i(x_1,...,x_{i+1})&=&\sum_{k=1}^{i} \frac{x_k - x_{i+1}}{\sqrt{i+i^2}} , 
\en
respectively. This particular choice of coordinates transforms the Hamiltonian into a set of $2N-2$ independent harmonic oscillators, without additional prefactors. The remaining two eigenvalues of $A$ are not degenerated, such that their corresponding eigenvectors are predefined, and imply the coordinate
\eq
\label{ChangeMassCenters1}
U_1 (R_{N_n},r_{N_e}) &=& \frac{N_n (a+b) R_{N_n}+N_e r_{N_e}}{\sqrt{N_e+N_n (a+b)^2}}, \\
\label{ChangeMassCenters2}
U_2 (R_{N_n},r_{N_e}) &=& \frac{N_n (a-b) R_{N_n}+N_e r_{N_e}}{\sqrt{N_e+N_n (a-b)^2}},
\en
where
\eq
\label{JacobiChangeCentermass}
r_{N_e}(x_1,...,x_{N_e})&=&\frac{1}{N_e}\sum_{i=1}^{N_e} x_i, \nonumber \\
R_{N_n}(X'_1,...,X'_{N_n})&=&\frac{1}{N_n}\sum_{j=1}^{N_n} X'_i,
\en
are the centers-of-mass of the electrons and the nuclei, respectively, and
\eq
\label{ChangeParameters}
a &=& \frac{M-1-2 \tau_{ne}+M \tau_{ne}}{2 \sqrt{M} \tau_{ne} }, \\
b &=&\frac{\sqrt{-4 M \left(1+3 \tau_{ne}\right)+\left(1+M+2\tau_{ne}+M \tau_{ne}\right)^2}}{2 \sqrt{M} \tau_{ne}}. \nonumber
\en

In the transformed coordinates $\{ r_1, \ldots, r_{N_e -1}\}$, $\{R_1, \ldots, R_{N_n - 1}\}$ and $\{U_1,U_2\}$, the Hamiltonian \eqref{ManyParticleHamiRescaled} separates as
\eq
\label{SepHamiltonian}
H'_x = 
\sum_{l=1,2}\left(-\frac{1}{2}\frac{\partial^2}{\partial U_l^2} + \frac{1}{2} \beta_l U_l^2 \right) + \nonumber \\
\sum_{j=1}^{N_n-1}\left(-\frac{1}{2}\frac{\partial^2}{\partial R_j^2} + \frac{1}{2} \beta^{(n)} R_j^2 \right) + \nonumber \\
\sum_{i=1}^{N_e-1}\left(-\frac{1}{2}\frac{\partial^2}{\partial r_i^2} + \frac{1}{2} \beta^{(e)} r_i^2 \right),
\en
where
\eq
\label{beta1beta2}
\beta_{1/2} & = &\frac{1+M + N_e \tau_{ne}+ N_n M \tau_{ne}}{2 M} \mp \frac{N_n b \tau_{ne}}{\sqrt{M}}  \\
\label{betan}
\beta^{(n)} & = & \frac{1+ N_e \tau_{ne}+N_n\tau_{nn}}{M}  \\
\label{betae}
\beta^{(e)} & = & 1+N_n\tau_{ne}+N_e\tau_{ee}.
\en
In other words, the system has been decomposed into a set of independent harmonic oscillators in the variables $U_1$, $U_2$, $R_j$ and $r_i$, with frequencies $\sqrt{\beta_1}$, $\sqrt{\beta_2}$, $\sqrt{\beta^{(n)}}$, $\sqrt{\beta^{(e)}}$, respectively. The eigenfunctions of the Hamiltonian
\eq
\label{ExactWaveFunctionIndepOsci}
\Psi'_{u_1,u_2,\mathbf{n},\mathbf{e}}(X'_1,...,X'_{N_n},x_1,...,x_{N_e}) = \nonumber \\
\prod_{j=1}^{N_n-1} \Phi_{n_j}^{\beta^{(n)}} \left(R_i\right)
\prod_{i=1}^{N_e-1} \Phi_{e_i}^{\beta^{(e)}} \left(r_i\right) \times \nonumber \\
\times \prod_{l=1}^2 \Phi_{u_l}^{\beta_l}\left(U_l(R_{N_n},r_{N_e})\right)
\en
are expressed in terms of the one-dimensional harmonic oscillator solution
\eq
\label{HarmonicWaveFunction}\label{harmonic-solution}
\Phi_{\nu}^{\beta}(y)= \left( \frac{\beta^{1/4}}{2^{\nu}\nu!\pi^{1/2}} \right)^\frac{1}{2} e^{-\frac{1}{2}\sqrt{\beta}y^2} \mathcal{H}_{\nu}\left( \beta^{1/4}y \right)
\en
where $\mathcal{H}_\nu(y)$ denotes the Hermite polynomial. The quantum numbers $u_1, u_2, \mathbf{n}, \mathbf{e}$ correspond to the excitation of each collective coordinate \eqref{ChangeMassCenters1}, \eqref{ChangeMassCenters2}, \eqref{JacobiChangeRelative2} and \eqref{JacobiChangeRelative1} respectively; that is, the quantum numbers $u_i$ are associated to the excitation of coordinates $U_i$, and $\mathbf{n}$ ($\mathbf{e}$) denotes the set of quantum numbers $\{n_1,...,n_{N_n-1}\}$ ($\{e_1,...,e_{N_e-1}\}$) associated to the excitation of the nuclei (electrons) relative coordinates $\{R_1,...,R_{N_n-1}\}$ ($\{r_1,...,r_{N_e-1}\}$).

The eigenfunctions of the initial Hamiltonian $H_x$ are obtained by undoing the dilatation coordinates change \eqref{dilatation} in the eigenfunction given in Eq. \eqref{ExactWaveFunctionIndepOsci}, {\it i.e.}
\eq
\label{1DExactWaveFunction}
\Psi_{u_1,u_2,\mathbf{n},\mathbf{e}}(\mathbf{X},\mathbf{x}) = M^{-\frac{N_n}{4}} 
\Psi'_{u_1,u_2,\mathbf{n},\mathbf{e}} \left(\sqrt{M}\mathbf{X},\mathbf{x}\right),
\en
where $\mathbf{x}$ ($\mathbf{X}$) is the set of the electron (nuclei) positions $\{{x}_i\}$ ($\{{X}_i\}$).
While the above eigenfunctions depend on the rescaled interactions between particles and the mass ratio, the eigenenergies depend explicitly on all parameters that govern the physical scale of the system, that is, the energy of the system, $E_x = \sqrt{\frac{k}{m_e}} E'_x$, is given by
\eq
\label{eigenenergies}
E_x &=& \sqrt{\frac{k}{m_e}} \left( \sum_{l=1}^2 \sqrt{\beta_l} \left( u_l + \frac{1}{2} \right) \right.+ \\
&+&\left. \sum_{j=1}^{N_n-1} \sqrt{\beta^{(n)}_j} \left( n_j + \frac{1}{2} \right) + \sum_{i=1}^{N_e-1} \sqrt{\beta^{(e)}_i} \left( e_i + \frac{1}{2} \right)\right) \nonumber,
\en
where $k$ is the strength of the confining potential and $m_e$ ($m_n$) is the electron (nucleus) mass such that $M=m_n/m_e$.

We denote pure states of the system \eqref{ManyParticleHamiltonian} by $|u_1,u_2,\mathbf{n},\mathbf{e}\rangle$, resulting in the wavefunction \eqref{1DExactWaveFunction}
\eq
\label{1DExactWaveFunctionState}
\Psi_{u_1,u_2,\mathbf{n},\mathbf{e}}(\mathbf{X},\mathbf{x}) = \langle \mathbf{X},\mathbf{x}|u_1,u_2,\mathbf{n},\mathbf{e}\rangle.
\en

\end{appendix}

% \bibliography{EntanglemetHarmonicSystems}
% \bibliographystyle{h-physrev5}

\end{document}